\documentclass[conference]{IEEEtran}
\IEEEoverridecommandlockouts

\usepackage{cite}
\usepackage{amsmath,amssymb,amsfonts}
\usepackage{algorithmic}
\usepackage{graphicx}
\usepackage{textcomp}
\usepackage{xcolor}
\usepackage{subfigure}
\usepackage{booktabs}
\usepackage{longtable} 
\usepackage{multirow}
\usepackage{colortbl}
\usepackage[colorlinks=true,pdfborder={0 0 0}]{hyperref}
\usepackage{fancyhdr}

\def\BibTeX{{\rm B\kern-.05em{\sc i\kern-.025em b}\kern-.08em
		T\kern-.1667em\lower.7ex\hbox{E}\kern-.125emX}}

\begin{document}

\title{RAHN: A Reputation Based Hourglass Network for Web Service QoS Prediction\thanks{DOI reference number: 10.18293/SEKE2024-024}
}

\author{
        \IEEEauthorblockN{
                Xia Chen\textsuperscript{1,2}, 
                Yugen Du\textsuperscript{1,2*},
                Guoxing Tang\textsuperscript{1,2}, 
                Yingwei Luo\textsuperscript{1,2} and
                Benchi Ma\textsuperscript{1,2}
            }
        \IEEEauthorblockA{
                \textsuperscript{1}\textit{Software Engineering Institute, East China Normal University, Shanghai 200062, China}\\
                \textsuperscript{2}\textit{Shanghai Key Laboratory of Trustworthy Computing, Shanghai 200062, China} \\
                \textit{Email: \{chenxia200062\}@gmail.com, \{ygdu\}@sei.ecnu.edu.cn}
            }
    }

\IEEEpeerreviewmaketitle 
\maketitle
\IEEEpeerreviewmaketitle 

\begin{abstract}
As the homogenization of Web services becomes more and more common, the difficulty of service recommendation is gradually increasing. How to predict Quality of Service (QoS) more efficiently and accurately becomes an important challenge for service recommendation. 
Considering the excellent role of reputation and deep learning (DL) techniques in the field of QoS prediction, we propose a reputation and DL based QoS prediction network, RAHN, which contains the Reputation Calculation Module (RCM), the Latent Feature Extraction Module (LFEM), and the QoS Prediction Hourglass Network (QPHN). 
RCM obtains the user reputation and the service reputation by using a clustering algorithm and a Logit model. 
LFEM extracts latent features from known information to form an initial latent feature vector. 
QPHN aggregates latent feature vectors with different scales by using Attention Mechanism, and can be stacked multiple times to obtain the final latent feature vector for prediction. We evaluate RAHN on a real QoS dataset. The experimental results show that the Mean Absolute Error (MAE) and Root Mean Square Error (RMSE) of RAHN are smaller than the six baseline methods.
\end{abstract}


\section{Introduction}
The number of Web services is increasing day by day. 
The emergence of homogenization causes the process of finding the most suitable Web service to become more difficult for users. 
In general, QoS are susceptible to network fluctuations, hardware failures, service errors, etc., and do not represent the service invocation experience of users under normal conditions \cite{tang2016collaborative}.
Therefore, if we want to recommend the most suitable Web service for each user, we need to obtain the corresponding QoS information in advance.
Since obtaining the QoS information corresponding to all the Web services requires a lot of time and resource costs. Thus, researchers consider QoS prediction as a key method to obtain QoS information and have conducted extensive and in-depth research on it.

Nowadays, MF and DL have been widely recognized as the most general model-based QoS prediction methods. 
Since the reliability of historical QoS information can have a significant impact on MF-based QoS prediction, the effectiveness of QoS information can be affected by unreliable users and services.
Therefore many researchers have tried to use reputation to quantify the reliability of users and services.
With the development and popularization of DL technology in CV, and NLP, some researchers also try to apply it to QoS prediction and confirm the importance of DL technology in QoS prediction.

After combining reputation and DL technology, this paper proposes an hourglass QoS prediction method based on reputation and DL, RAHN. 
Due to the arbitrariness of some users in providing QoS observations and the inherent instability of some Web services, it is necessary to take the user reputation and service reputation into consideration.

In summary, the main contributions of this paper can be summarized as follows: 
\textbf{1)} We propose a QoS prediction network RAHN, which contains RCM, LFEM, and QPHN. 
\textbf{2)} RCM can calculate user reputation and service reputation, which makes RAHN better robust to unreliable users and unreliable Web services. 
\textbf{3)} LFEM can extract latent features from the information and reputation of users and Web services. 
\textbf{4)} QPHN can mine high-level features and aggregate latent features at different scales to build better latent feature vector. 
\textbf{5)} Multiple experiments were conducted on the dataset containing a large amount of real QoS data to verify the effectiveness and superiority of RAHN.

\section{Related Work}
\subsection{MF-based QoS prediction}
MF is a method that can fully utilize sparse QoS information for QoS prediction. Since its development, many researchers have proposed their own thinking for it. 
Zhong et al. \cite{zhong2022collaborative} proposed Network Bias MF (NBMF) considering that different users may have different degrees of delay when invoking Web services, and achieved better prediction performance.
Chen et al. \cite{chen2022web} used Dirichlet distribution to compute the user reputation and achieved ideal results.

These works have demonstrated the important support of reliability for QoS prediction. Therefore, this paper incorporates both user reliability and service reliability into the thinking and calculates user and service reputation to assist QoS prediction.

\subsection{DL-based QoS prediction}
Generally speaking, DL models can utilize complex user information and Web service information more effectively to improve QoS prediction accuracy. 
Peng et al. \cite{peng2023th} proposed a web service link prediction method based on topic-aware heterogeneous graph neural networks. It captures fine-grained topic-aware semantics while mining contextual topic distribution to achieve better prediction results. 
Jia et al. \cite{jia2022location} overcomes QoS data sparsity by fusing local and global location information of users and Web services in the interaction layer, using MLP to obtain high-dimensional nonlinear relationships and combining them with low-dimensional linear relationships.

All of the above works have shown that DL techniques have great potential in QoS prediction tasks and can lead to even better QoS prediction performance. Inspired by them, this paper designs RCM, DL based LFEM and DL based QPHN to improve the QoS prediction accuracy.

\section{Preliminaries}
\begin{figure}[ht]
    \centering
    \includegraphics[width=0.8\linewidth,height=0.4\linewidth]{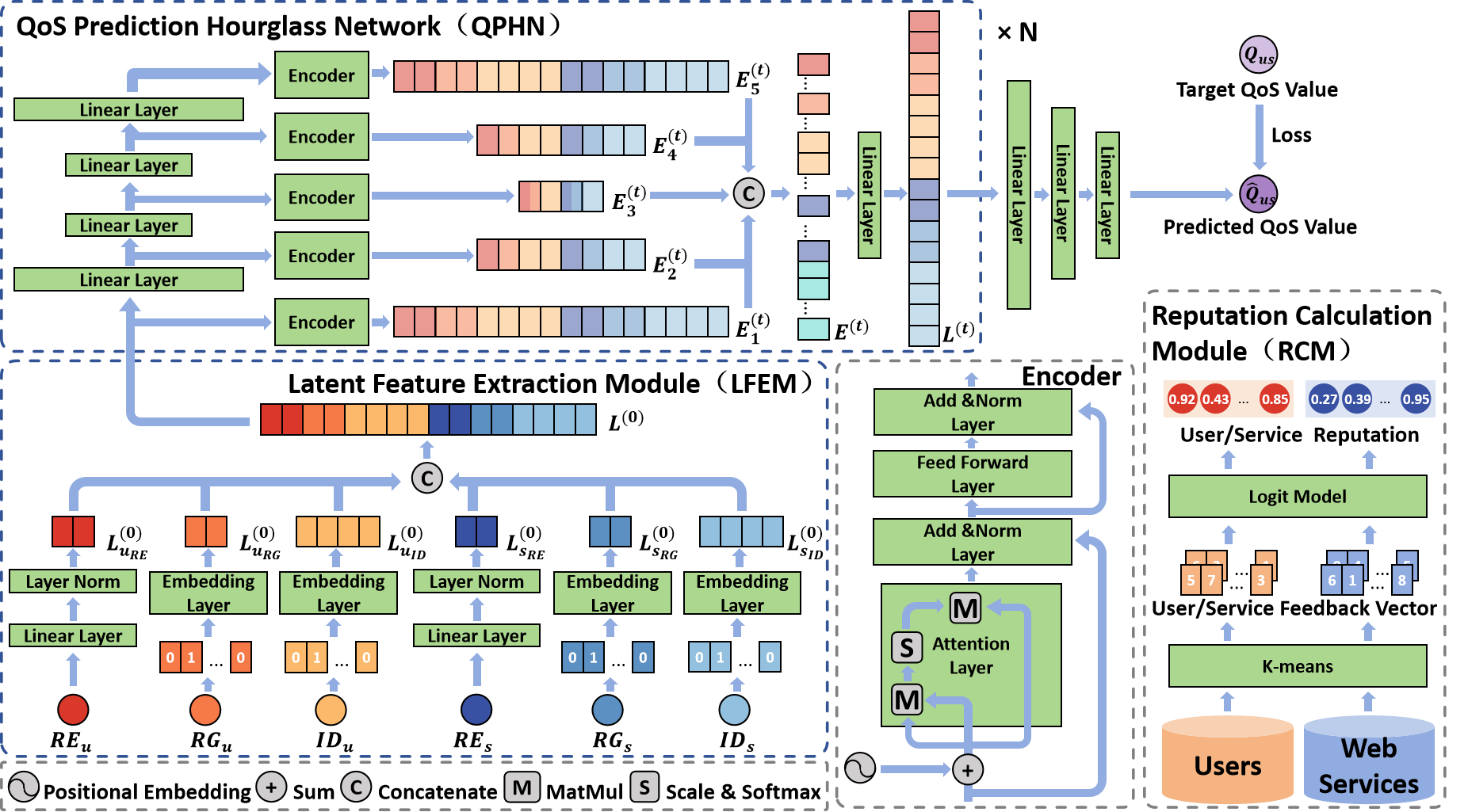}
    \caption{The Overall Architecture of RAHN.
    1) \textbf{RCM:} We use the K-means algorithm to cluster users and Web services into clusters, and then count the user feedback vectors and service feedback vectors according to the $3\sigma$ rule in the normal distribution. Then we apply Logit model to calculate user reputation and service reputation.
    2) \textbf{LFEM:} We use Linear Layers and Embedding Layers to extract user latent features from user information and user reputation. And in the same way we extract Web service latent features. Then, user latent features and service latent features are concatenated together to form the final latent feature vector.
    3) \textbf{QPHN:} We process the latent feature vectors using Linear Layers and Encoders at different scales to obtain multi-scale feature vectors with high-level feature information. These latent feature vectors are then concatenated and multiple QPHNs can be stacked for use.
    }\label{fig:01}
\end{figure}
\subsection{Problem Formulation}
In this section some necessary definitions will be given:
\begin{itemize}
    \item \textbf{User Reputation (in Eq.\eqref{eq:11}):}
    Certain users are casual and heavily subjective in submitting QoS observations. User reputation measures how trustworthy a user is.
    \item \textbf{Service Reputation (in Eq.\eqref{eq:11}):}
    Some Web services may have QoS fluctuations due to their own bugs. Service reputation measures how stable a Web service is in providing QoS.
    \item \textbf{NPEd:} Parameter combinations in the experiment, e.g. NPEd=0108 denotes N=0, PE=1, d=08. It will be mentioned in the Fig.\ref{fig:02} and \ref{fig:04} in section V.
\end{itemize}

\subsection{Method Overview}
Fig.\ref{fig:01} illustrates the overview architecture of RAHN and individual modules. The details of these modules will be shown in Section IV.

\section{Proposed Approach}
\subsection{RCM}
RCM can calculate the user reputation and service reputation. We use the K-means clustering algorithm to cluster users and Web services separately to obtain $N_u$ user clusters and $N_s$ service clusters. According to \cite{wu2015qos}, we consider the cluster containing the most elements as the reliable cluster. The specific definition formulas are as follows:
$U_r=\{u|u\in C_{u}^{x},x=\mathop{argmax}\limits_{k}\vert C_{u}^{k}\vert,1\leq k\leq N_u\}$,
$S_r=\{s|s\in C_{s}^{x},x=\mathop{argmax}\limits_{k}\vert C_{s}^{k}\vert,1\leq k\leq N_s\}$,
where $U_r$ denotes the set of reliable users, $C_{u}^{k}$ denotes the $k$-th user cluster, $N_u$ denotes the total number of user clusters, $S_r$ denotes the set of reliable services, $C_{s}^{k}$ denotes the $k$-th service cluster, $N_s$ denotes the total number of service clusters, and $|C^{k}|$ denotes the number of elements in the $k$-th cluster.

Based on the work of \cite{su2016web}, we believe that reliable clusters reflect what the public perceives as normal QoS observations, and that the QoS values should follow a normal distribution $N(\mu,\sigma^{2})$ ($\mu$ is the mean and $\sigma$ is the standard deviation). Based on the reliability clusters, we classify all QoS observations into two types: positive and negative feedback. QoS observations that are close to the normal value are positive feedback and vice versa. According to the $3\sigma$ rule, we consider QoS observations that are within the interval $(\mu_{r}-3\sigma_{r},\mu_{r}+3\sigma_{r})$ to be positive feedback, and vice versa to be negative feedback. Both $\mu_{r}$ and $\sigma_{r}$ come from reliable clusters. The feedback vector can then be obtained from $F=[po,ne]$,
where $po$ indicates the amount of positive feedback and $ne$ indicates the amount of negative feedback.

Suppose, the probability and utility of the user providing positive and negative feedback are $p_1$, $p_2$, $U_1$, $U_2$, as shown in Eq.\eqref{eq:05}. 
According to the principle of utility maximization in economics, the user will choose the option with greater utility. 
Thus $p_1$, $p_2$ are defined as follows:
\begin{align}
    U_{1}&=V_{1}+\epsilon_{1}, 
    U_{2}=V_{2}+\epsilon_{2}, \label{eq:05}\\
    p_{1}&=P(U_{1}>U_{2}),
    p_{2}=P(U_{1}<U_{2}), \label{eq:07}
\end{align}
where the $V_1$ and $V_2$ denote the observable deterministic parts, and $\epsilon_1$ and $\epsilon_2$ denote the random terms.

Assume that $\epsilon_1$ and $\epsilon_2$ follow the standard Gumbel distribution and are independent of each other. Then $\epsilon_1-\epsilon_2$ will satisfy the $Logistic(0,1)$. Its distribution function is as follows:
\begin{align}    
    F(x)&=P(X\leq x)
    =\frac{1}{1+e^{-(x-\mu)/\gamma}},(\mu=0,\gamma=1).\label{eq:08}
\end{align}
Assuming $V = \beta X$, then $p_{1}$ and $p_{2}$ can be transformed into:
\begin{align}
    p_{1}&=F(V_{1}-V_{2})
    =\frac{e^{\beta X_{1}}}{e^{\beta X_{1}}+e^{\beta X_{2}}}=\frac{e^{\beta po}}{e^{\beta po}+e^{\beta ne}}, \label{eq:09}\\
    p_{2}&=F(V_{2}-V_{1})
    =\frac{e^{\beta X_{2}}}{e^{\beta X_{1}}+e^{\beta X_{2}}}=\frac{e^{\beta ne}}{e^{\beta po}+e^{\beta ne}}, \label{eq:10}
\end{align}
where $\beta$ is the positive coefficient of $X$.
Finally, reputation can be calculated from both positive and negative feedback:
\begin{align}
    Re={p_{1}}/{(p_{1}+p_{2})},\label{eq:11}
\end{align}
where Re denotes reputation and the value is in the range [0,1].

\subsection{LFEM}
LFEM can obtain an initial latent feature vector $L^{(0)}$. Each user or Web service has its own information (ID, region (RG) and previously obtained reputation). We use Linear Layer to extract the hidden information in the reputation and obtain the feature vectors $L_{u_{RE}}^{(0)}$, $L_{s_{RE}}^{(0)}$. Then we transform ID and RG into one-hot vectors \cite{yin2020qos} respectively, and obtain the links in them by Embedding Layer to generate feature vectors $L_{u_{RG}}^{(0)}$, $L_{u_{ID}}^{(0)}$, $L_{s_{RG}}^{(0)}$, $L_{s_{ID}}^{(0)}$. These feature vectors will be concatenated to obtain $L^{(0)}$, which is represented as follows:
\begin{align}
    L^{(0)}=L_{u_{RE}}^{(0)}\oplus L_{u_{ID}}^{(0)}\oplus L_{u_{RG}}^{(0)}\oplus
    L_{s_{RE}}^{(0)}\oplus L_{s_{ID}}^{(0)}\oplus L_{s_{RG}}^{(0)},
\label{eq:12}\end{align}
where $\oplus$ denotes the concatenate operation,
$L_{{RE}}^{(0)}$$\in$$\mathbb{R}^{1\times (d/4)}$, 
$L_{{RG}}^{(0)}$$\in$$\mathbb{R}^{1\times (d/2)}$, 
$L_{{ID}}^{(0)}$$\in$$\mathbb{R}^{1\times (d/4)}$
and $d$ is a positive integer divisible by 4.
$L^{(0)}$$\in$$\mathbb{R}^{1\times 2d}$, which blends the user with the latent features in the Web service.

\subsection{QPHN}
QPHNs can obtain latent feature vectors at multiple scales, and can be repeatedly stacked.
We take $L^{(0)}$ as the initial input to the QPHNs and consider the processing of a QPHN as a function $\Phi(x)$ for obtaining the final latent feature vector $L^{(n)}$. The formula is shown below:
\begin{align}
    L^{(n)}=\Phi_{n}(\Phi_{n-1}...(\Phi_{1}(L^{(0)}))),
\label{eq:13}\end{align}
where $n>0$ denotes the number of stacked layers of the QPHN, and by $n$ iterations, we transform  $L^{(0)}$ to $L^{(n)}$.

For the $t$-th layer QPHN, its input is $L^{(t-1)}$. We use Linear Layers of different sizes to obtain multi-scale latent feature vectors:
\begin{align}
    L_{1}^{(t)}=L^{(t-1)},
    L_{k}^{(t)}=ReLU(g_{k-1}(L_{k-1}^{(t)})),
\label{eq:15}\end{align}
where $L_{1}^{(t)}$$\in$$\mathbb{R}^{1\times 2d}$, $L_{2}^{(t)}$$\in$$\mathbb{R}^{1\times d}$, 
$L_{3}^{(t)}$$\in$$\mathbb{R}^{1\times (d/2)}$, 
$L_{4}^{(t)}$$\in$$\mathbb{R}^{1\times d}$, 
$L_{5}^{(t)}$$\in$$\mathbb{R}^{1\times 2d}$, 
denote the latent feature vectors of different scales, respectively. 
k=2,3,4,5.
ReLU \cite{glorot2011deep} is the activation function that we use. $f^{[a,b]}(x)$ denotes the linear layer, with $a$ being the dimension of the input $x$ and $b$ being the output dimension. 
$g_{1}$=$f^{[2d,d]}$,$g_{2}$=$f^{[d,d/2]}$,$g_{3}$=$f^{[d/2,d]}$,$g_{4}$=$f^{[d,2d]}$.

Then, Encoders with different scales are used to extract high-level features in $L_{i}^{(t)}$ and form the new latent feature vector $E_{i}^{(t)}=Encoder(L_{i}^{(t)})$,
where the structure of Encoder is shown in Fig.\ref{fig:01} and $i\in[1,5]$. We utilize the attention mechanism to focus on the key information in $L_{i}^{(t)}$ at different scales to obtain a more effective latent feature vector $E_{i}^{(t)}$.
Next, we concatenate $E_{i}^{(t)}$ to get $E^{(t)}$ and adjust its dimension to obtain the same size as the input $L^{(t-1)}$ to get $L^{(t)}$:
\begin{align}
    &E^{(t)}=E_{1}^{(t)}\oplus E_{2}^{(t)}\oplus E_{3}^{(t)}\oplus E_{4}^{(t)}\oplus E_{5}^{(t)}, \label{eq:17}\\
    &L^{(t)}=f^{[13d/2,2d]}(E^{(t)}),
\label{eq:18}\end{align}
where $E^{(t)}$$\in$$\mathbb{R}^{1\times (13d/2)}$ aggregates multi-scale latent feature vectors, containing both low-level feature information and high-level feature information.

\subsection{QoS Prediction and Model Training}
With the previous modules we obtained the final latent feature vector $L^{(n)}$. We get the predicted QoS value $\widehat{Q}_{us}$ by three Linear Layers which can be represented as:
$\widehat{Q}_{us}=f^{[d/2,1]}(ReLU(f^{[d,d/2]}(ReLU(f^{[2d,d]}(L^{(n)})))))$,
where $\widehat{Q}_{us}$ denotes the QoS prediction result when user $u$ invokes Web service $s$.

RAHN is trained by loss function, and our goal is to minimize the loss function J, which can be expressed as:
$J=\frac{1}{N}\sum\limits_{k=1}^{N}\vert Pred(x_k,\Theta)-Q_{x_k}\vert+\lambda\mathbb{L}_{reg}(\Theta)$,
where N is the batch size, $x_k$ denotes the $k$-th input (containing information about the corresponding user and Web service), $\Theta$ denotes all the parameters to be learned, $Pred$ denotes the function of RAHN that maps the input $x_k$ to the predicted QoS value, and $Q_{x_k}$ denotes the target QoS value. In addition, $\lambda\mathbb{L}_{reg}(\Theta)$ is the regularization term, which is used to prevent the model from overfitting.

RAHN uses the Adam optimizer \cite{kingma2014adam} to update all parameters to be optimized as follows:
$\Theta\leftarrow\Theta -\eta\frac{\partial{J}}{\partial{\Theta}}$,
where $\eta$ is the learning rate, which is used to control the rate at which the model converges.

\section{Experiments}
\subsection{Dataset}
This paper conducts extensive experiments on the WS-DREAM \cite{zheng2010distributed} dataset with a large amount of real-world QoS data. WS-DREAM dataset contains 339 users, 5,825 Web services, and 2 user-service matrices (response time matrix and throughput matrix). In this paper, we use the user-service matrix of response time to evaluate the performance of RAHN and compare it with other baseline methods.

\subsection{Evaluation Metrics}
In this paper, MAE and RMSE are used as evaluation metrics. The related definitions are as follows:
$MAE=\frac{\sum\limits{\vert Q_{ij}-\widehat{Q}_{ij}\vert}}{N}$,
$RMSE=\sqrt{\frac{\sum\limits{(Q_{ij}-\widehat{Q}_{ij})^2}}{N}}$,
where $Q_{ij}$ denotes the true value, $\widehat{Q}_{ij}$ denotes the predicted value, and N denotes the number of entries in the test dataset.
Smaller MAE and RMSE indicate higher QoS prediction accuracy.

\subsection{Performance Comparison}
\begin{table}[h]
\renewcommand\arraystretch{1.5}
\centering
\caption{Performance Comparison of Response Time (Smaller MAE and RMSE Means Higher Accuracy)}
\scalebox{0.42}{
\begin{tabular}{ccccccccccccc}
\hline
\multirow{2}{*}{\textbf{Methods}} & \multicolumn{5}{c}{\textbf{MAE}}                                                         & \textbf{}        & \multicolumn{5}{c}{\textbf{RMSE}}                                                        & \textbf{}        \\ \cline{2-13} 
                                  & \textbf{MD=2\%} & \textbf{MD=4\%} & \textbf{MD=6\%} & \textbf{MD=8\%} & \textbf{MD=10\%} & \textbf{Improve} & \textbf{MD=2\%} & \textbf{MD=4\%} & \textbf{MD=6\%} & \textbf{MD=8\%} & \textbf{MD=10\%} & \textbf{Improve} \\ \hline
PMF \cite{mnih2007probabilistic}                               & 0.327           & 0.255           & 0.242           & 0.239           & 0.238            & 41.73\%          & 0.529           & 0.461           & 0.460           & 0.457           & 0.454            & 22.20\%          \\
CMF \cite{ye2021outlier}                               & 0.223           & 0.203           & 0.197           & 0.161           & 0.150            & 25.10\%          & 0.412           & 0.395           & 0.386           & 0.346           & 0.339            & 6.33\%           \\
RLMF \cite{chen2022web}                              & 0.185           & 0.162           & 0.162           & 0.156           & 0.147            & 16.99\%          & 0.391           & 0.370           & 0.368           & 0.359           & 0.344            & 4.65\%           \\
NeuMF \cite{he2017neural}                             & 0.163           & 0.150           & 0.148           & 0.149           & 0.147            & 12.18\%          & 0.366           & 0.358           & 0.356           & 0.354           & 0.353            & 2.72\%           \\
HSA-Net \cite{wang2021hsa}                           & 0.182           & 0.159           & 0.128           & 0.128           & 0.126            & 8.15\%           & 0.558           & 0.495           & 0.470           & 0.448           & 0.442            & 23.35\%          \\
PLRes \cite{zhang2021probability}                             & 0.174           & 0.143           & 0.135           & 0.125           & 0.122            & 5.90\%           & 0.524           & 0.461           & 0.424           & 0.409           & 0.407            & 18.18\%          \\ \hline
\textbf{RAHN}                     & \textbf{0.156}  & \textbf{0.134}  & \textbf{0.125}  & \textbf{0.118}  & \textbf{0.115}   & \textbf{-}       & \textbf{0.366}  & \textbf{0.348}  & \textbf{0.343}  & \textbf{0.337}  & \textbf{0.335}   & \textbf{-}       \\ \hline
\end{tabular}}\label{table:01}
\end{table}
To demonstrate the effectiveness of RAHN, we compare it with six other representative baseline methods. The details are presented as shown below:
\textbf{1) PMF:} This MF-based method incorporates probability.
\textbf{2) CMF:} This MF-based approach takes outliers into consideration.
\textbf{3) RLMF:} This MF-based method calculates the user reputation using the Dirichlet distribution.
\textbf{4) NeuMF:} This DL-based method fuses MF with multi-layer perceptron.
\textbf{5) HSA-Net:} This DL-based method uses hidden state awareness techniques.
\textbf{6) PLRes:} This DL-based method reutilizes the historical call probability distribution of users and services with location features.

We initialized the parameters of every baseline method according to the their paper to obtain the best performance.  
The parameters of RAHN are set to 
$N_u$=5, 
$N_s$=15, 
$N$=2 (Same as n in Eq.\eqref{eq:13}),
$PE$=0 (whether to use Position Embedding, 0 is false, 1 is true), 
$d$=16 (in Eq.\eqref{eq:12}).

We randomly remove the QoS data from the user-service matrix as a training matrix to study the prediction performances under different Matrix Densities (MD).
A test matrix is then created using the deleted data. MD is set to \{2\%, 4\%, 6\%, 8\%, 10\%\}. 
Outliers inevitably exist in historical QoS data, and if the model's predictions coincide with these outliers, this may lead to a situation where MAE and RMSE are small but the prediction accuracy is poor. Therefore, we exclude some outliers when calculating MAE and RMSE. 
In this paper, we removed 10\% of the most obvious outliers in the test data for all methods according to the steps in \cite{ye2021outlier}.
As shown in Table \ref{table:01}, when the matrix density is set in the range of 2\% to 10\%, RAHN has better prediction accuracy.

\subsection{Impact of Parameters}

\begin{figure}[h]
    \centering
    \subfigure{%
        \includegraphics[width=0.18\textwidth, height=0.15\linewidth]{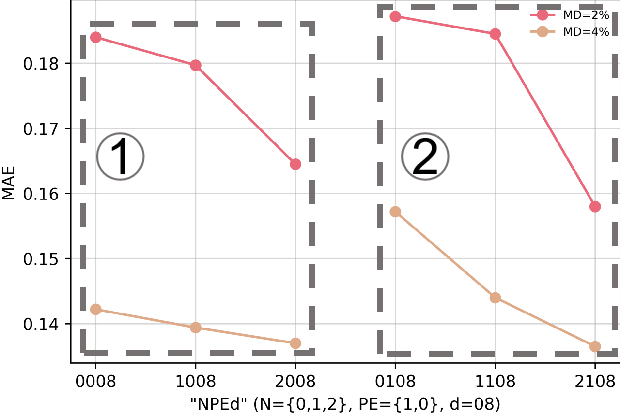}%
    }
    \quad
    \subfigure{%
        \includegraphics[width=0.18\textwidth, height=0.15\linewidth]{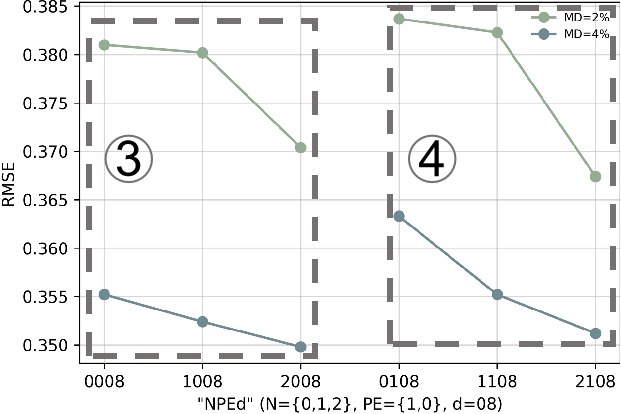}%
    }
    \caption{Impact of N}\label{fig:02}
    \subfigure{%
        \includegraphics[width=0.18\textwidth, height=0.15\linewidth]{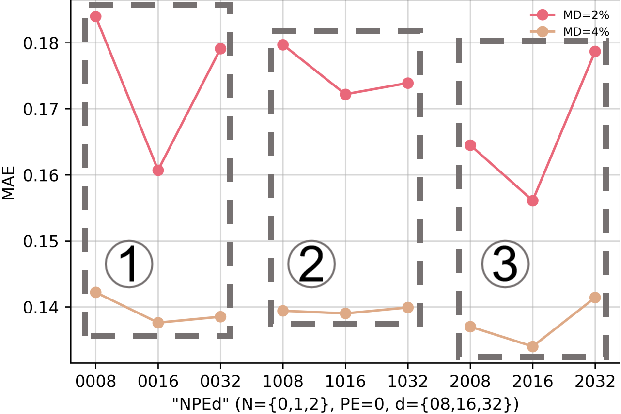}%
    }
    \quad
    \subfigure{%
        \includegraphics[width=0.18\textwidth, height=0.15\linewidth]{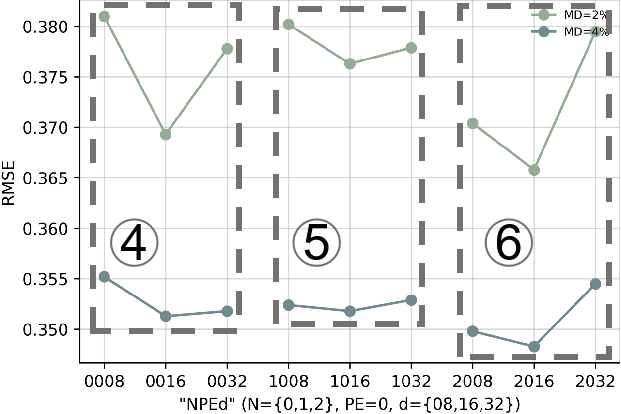}%
    }
    \caption{Impact of d}\label{fig:04}
\end{figure}
In order to analyze how different parameter settings affect the QoS prediction of RAHN, in this paper, the same parameter settings except the parameters to be analyzed are as follows: $N_u$=5, $N_s$=15, $\eta$=0.0005. MD is set to 2\% and 4\%.

\subsubsection{\textbf{Impact of N}}
N denotes the number of QPHN. The size of N determines the network depth of RAHN. In order to explore the effect of N on the prediction results, we conducted experiments with N set to \{0,1,2\} for PE=\{1,0\} and d=08, respectively. As shown in Fig. \ref{fig:02}, we can find from regions \normalsize{\textcircled{\scriptsize{1}}} - \normalsize{\textcircled{\scriptsize{4}}} that the MAE and RMSE decrease with the increase of N. Considering the time and resource costs, we believe that RAHN predicts best at N = 2.

\subsubsection{\textbf{Impact of d}}
d denotes the user/service latent feature dimension. In order to explore the effect of d on the prediction results, we conducted experiments with d set to \{08,16,32\} in the case of N=\{0,1,2\} and PE=0, respectively. As shown in Fig. \ref{fig:04}, we can find that both MAE and RMSE reach the minimum value at d=16 from regions \normalsize{\textcircled{\scriptsize{1}}} - \normalsize{\textcircled{\scriptsize{6}}}. Therefore, we believe that RAHN has the best prediction accuracy at d = 16.

\section{Conclusion}
Previously, little or no work has been done to integrate reputation with DL techniques, while we propose a reputation and DL based QoS prediction method, RAHN, with good predicted results. RAHN contains three modules:
RCM, LFEM, and QPHN. RCM calculates user reputation and service reputation; LFEM extracts the initial latent feature vectors from reputation, region, and ID; QPHN extracts the initial latent feature vectors that incorporate different scales.
To evaluate the effectiveness of RAHN, we conduct extensive experiments on real large-scale QoS datasets. Compared with the six baseline methods, RAHN reduces the MAE by an average of $5.9\% \sim 41.73\%$ and the RMSE by an average of $2.72\% \sim 23.35\%$.
In future, we plan to extend RAHN as a time-aware QoS prediction based method.

\bibliographystyle{elsarticle-num}
\bibliography{references}

\end{document}